\documentclass[useAMS,usenatbib,usegraphicx]{mn2e}

\usepackage{amssymb}
\usepackage{textcomp}

\title[Ultra-fine dark matter structure in the Solar neighbourhood]{Ultra-fine dark matter structure in the Solar neighbourhood}
\author[Fantin et al.]
{Daniele S. M.\ Fantin$^1,^2$\thanks{E-mail: daniele.fanta21@gmail.com},
Anne M. Green$^1$,
Michael R. Merrifield$^1$\\
$^1$School of Physics \& Astronomy, University of Nottingham, 
    University Park, Nottingham, NG7 2RD\\
$^2$Centro de Investigaciones de Astronom\'{i}a, 
    Apdo. Postal 264, M\'{e}rida 5101-A, Venezuela\\
}
\begin{document}

\newcommand{\apj} {ApJ}
\newcommand{\aj} {AJ}
\newcommand{\apjl} {ApJL}
\newcommand{\aap} {AAp}
\newcommand{\apjs} {ApJS}
\newcommand{\aaps} {AApS}
\newcommand{\mnras} {MNRAS}
\newcommand{\pasp} {PASP}
\newcommand{\physrep} {Physics Reports}
\newcommand{\prd} {Physical Review D}
\newcommand{\nat} {Nature}


\pagerange{\pageref{firstpage}--\pageref{lastpage}} \pubyear{2011}

\maketitle

\label{firstpage}

\begin{abstract}

The direct detection of dark matter on Earth depends crucially on its density and its velocity distribution on a milliparsec scale. 
Conventional N-body simulations are unable to access this scale, making the development of other approaches necessary. 
In this paper, we apply the method developed in \cite{Fantin08} to a cosmologically-based merger tree, transforming it into a useful instrument to reproduce and analyse the merger history of a Milky Way-like system.
The aim of the model is to investigate the implications of any ultrafine structure for the current and next generation of directional dark matter detectors.
We find that the velocity distribution of a Milky Way-like Galaxy is almost smooth, due to the overlap of many streams of particles generated by multiple mergers.
Only the merger of a $10^{10} \, M_{\odot}$ analyse can generate significant features in the ultra-local velocity distribution, detectable at the resolution attainable by current experiments.

\end{abstract}


\section{Introduction}

In the last forty years, dark matter (hereafter DM) has been one of the more challenging topics in astronomy. 
After the first pioneering studies (Zwicky 1933 and Smith 1936), convincing evidence for the existence of DM came from the rotation curves of spiral galaxies (Freeman 1970, Einasto 1974, Rubin \& Ford 1970, Rubin et al. 1978, 1980, 1982, 1985). 
Its presence has been also supported by recent indirected measurements of its density, obtained by surveys such as SDSS (Tegmark et al. 2006), or by satellites like WMAP (Komatsu et al. 2009, Larson et al. 2010). 
Despite all this evidence, the nature of DM is still unknown. 
The most widely accepted idea is that it consists of non-relativistic, weakly interacting, non-baryonic particles (Cowsik \& McClelland 1973; Szalay \& Marx 1976), emitting no (or very little) electromagnetic radiation. 

Numerical simulations provide a reliable method of calculating the DM distribution on large scales, making robust predictions for its clustering. 
The resolution of the first simulations (Peebles 1982, Frenk et al. 1985) was not sufficient to resolve the inner parts of the DM halos, but the gap has started to be filled in the last decade, with N-body simulations able to provide results for halos of the size of the largest galaxy clusters (Springel et al. 2005, Boylan-Kolchin et al. 2009).
More recently simulations such as Via Lactea II (Diemand et al. 2008), GHALO (Stadel et al. 2009) and Aquarius (Springel et al. 2008) were able to reproduce the formation and the structure of a Milky Way-like DM halo down to a resolution of about 100 pc.

Particle physics provides us with various well-motivated dark matter candidates, including weakly interacting massive particles (hereafter WIMPs). 
They interact with ordinary matter through elastic scattering on atomic nuclei (Goodman \& Witten 1985), and many experiments are currently underway aiming to detect this phenomenon, such as CDMS II (Ahmed et al. 2010), Xenon100 (Aprile et al. 2010) and Zeplin III (Akimov et al. 2010). 
The goal of these experiments is to measure the number of recoil events per unit energy and, in some case, its temporal or angular dependence. 
These quantities depend on the local DM density and speed distribution of the incident particles in the detector rest frame (Jungman et al. 1996, Lewin \& Smith 1996). 
Therefore, the possible presence of a significant amount of fine-grained structure in the DM distribution in the Solar neighbourhood, such as overdensities and streams, would have important consequences for these direct detection experiments.
\\
The establishment of a reliable model for the Milky Way still represents a significant source of uncertainty.
One of the main reasons is that the scale relevant for direct detection experiments ($\sim 0.1 \rm{mpc}$) is well beyond the resolution of current simulations. 
Furthermore, DM particles are ``cold'', allowing the formation of micro-halos with masses down to $10^{-6} M_{\odot}$ (Green et al. 2004). 
Unfortunately, such micro-halos are far smaller than the smallest subhalos resolvable by simulations, which have mass $\sim 10^4 M_{\odot}$ and size $\sim 100 \rm{pc}$.
Consequently, other approaches are required to resolve the ultra-small scales probed by direct detection experiments. 
\\
Stiff and Widrow (hereafter SW) calculated the DM ultra-local velocity distribution $f(v)$ at a single spatial point of phase-space, identified as an ideal terrestrial detector (Stiff \& Widrow 2003). 
This result was achieved using a reverse technique that allowed them to reach a resolution that cannot be obtained by N-body simulations. 
They found that $f(v)$ is characterised by the presence of a small number of discrete streams of particles. 
Unfortunately, the reverse integration is unstable, requiring the introduction of a softening length of $20\, {\rm kpc}$ into the adopted gravitational force law. 
This softening is large enough to affect the DM phase-space distribution imprinting on a terrestrial detector.
\\
Other theoretical investigations have argued that the ultra-local WIMP distribution consists of a much larger number of streams, of the order of $10^{5-6}$ (Helmi et al. 2002; Vogelsberger et al. 2009). 
Overlapping DM streams in the inner part of the halo lead to more complex configurations, which may ultimately be indistinguishable from a smooth distribution. 
Moreover, direct cosmological simulations show that the most massive local streams only contribute about 1 percent of the local DM density (Diemand et al. 2008, Vogelsberger et al. 2011).
Schneider et al. (2010) found that at Earth's location the DM ultra-fine distribution is composed of over $10^3$ streams, which are generated by the tidal disruption of DM ``cold'' microhalos (Schneider et al. 2010). 
They conclude that the material these streams are composed of is not dense enough to be relevant for detection experiments.
\\
Recently Vogelsberger and White (2011; hereafter VW) have argued that the picture on ultra-local scales is not too different from that suggested by N-body simulations.
Using a new technique for calculating the phase-space distribution function in the neighbourhood of a simulation particle (Vogelsberger et al. 2008), they estimate the presence of about $\sim 10^{14}$ unresolved streams at the Solar position. 
Their conclusion is that the local velocity distribution is smooth (Vogelsberger \& White 2011).

In this paper, we present a new method of simulating the DM distribution of the Milky Way, building on the approach developed by Fantin et al. (2008; hereafter Paper I). 
Applied to a merger tree describing the history of a Milky Way-like galaxy, the model attributes to each progenitor of the halo a characteristic DM distribution, using the method developed in Paper I. 
The final result, obtained by summing the contributions of all the progenitors, provides a model of the ultra-fine DM distribution in the Solar neighbourhood at arbitrarily high spatial resolution, obtained without the computational overhead of a complete numerical integration.

The paper is organised as follows. 
In Section 2 we present the technique to calculate the ultra-fine DM distribution of a merger tree. 
Section 3 contains our results, Section 4 draws some conclusions, and Section 5 provides a summary.

\section{The model}

\begin{figure} 
\includegraphics[height=0.30\textheight, width=0.48\textwidth]{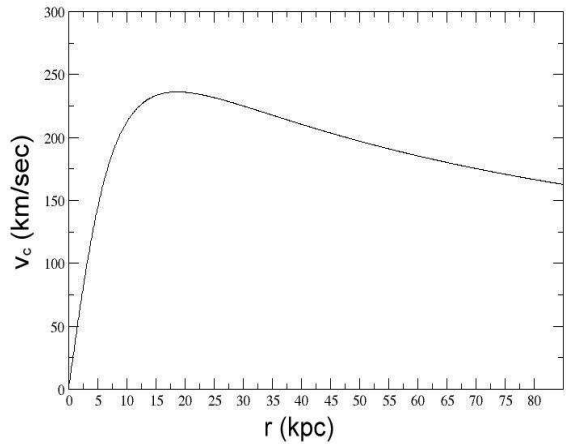}
\caption{The rotation curve of the isochrone potential, with the scale length taken to be equal to the Solar radius. 
Although not intended as a fully realistic model, it is a reasonable approximation to the measured rotation curve of the Milky Way (Fig. 5 of Sofue et al. 2009).}
\label{fig:rot_curve_isoc}
\end{figure}

In Paper I we modelled the interaction between an unbound system of particles and a Milky Way-like galaxy. 
Although not intended as a quantitative description of the Milky Way, this approach makes possible a detailed exploration of phase-space, and a thorough analysis of the expected signature of a series of merger events in a terrestrial DM detector. 
The method has the great benefit of being analytically soluble, allowing the calculation of the dynamics of the merger remarkably simply. 
Furthermore, the gravitational force does not have to be artificially softened, providing a significant improvement in accuracy.
To describe the Milky Way gravitational well we adopt the isochrone potential. 
Despite not being a realistic representation of a complex system such as the Milky Way, for example causing an overestimation of the granularity of the system at the present time, the qualitative properties of the potential are similar to those of our galaxy. 
Fig \ref{fig:rot_curve_isoc} presents the rotation curve calculated with a scale length equal to the solar radius.
Both the value of the circular speed at the Solar radius and the overall shape of the rotation curve are in reasonable agreement with the Milky Way rotation curve shown in Fig. 1 and 5 of Sofue et al. (2009)
%
%

Concerning the merging halo, we model its initial DM phase-space distribution by assuming a bivariate Gaussian, 
\begin{equation} \label{eq:disrt-func}
f(\mathbf{r},\mathbf{v})\propto \, e^{-[{(\mathbf{r}-\mathbf{r_{sat}})^2}/{2\sigma_{\rm{s}}^2})]} \; 
e^{-[({\mathbf{v}-\mathbf{v_{sat}})^2}/{2\sigma_{\rm{v}}^{2}}]} \ ,
\end{equation}
where ($\mathbf{r_{\rm{sat}}}$, $\mathbf{v_{\rm{sat}}}$) is the initial location of the merging halo in phase-space, which allows us to simply parametrise its main properties, where $\sigma_{\rm{s}}$ models its initial spatial extent and $\sigma_{\rm{v}}$ its velocity dispersion. 

Galaxy formation models (Kauffmann et al. 1993; Cole et al. 1994) and N-body simulations (Springel et al. 2005, Boylan-Kolchin et al. 2009) both suggest that galaxies formed though a continuous and regular hierarchical process. 
We therefore embed the model for a single merger into a cosmological context by combining it with a merger tree. 
This approach allows us to describe a system in which lots of DM satellites fall into the Galactic potential well.
Furthermore, it produces a more realistic estimate of the final velocity distribution of the Galactic halo than the one presented in Paper I because the merger tree describes a halo of $10^{12} M_{\odot}$, which has not undergone any recent major merger, as in the Milky Way.
The tree has been kindly generated by Andrew Benson, using a semi-analytic method (Cole et al. 2000, Parkinson et al. 2008) based on the assumption of the following cosmological parameters:  $\Omega_{0} = 0.25$, $\Lambda_{0}= 0.75$, $\Omega_{b} = 0.045$, $h_{0} = 0.73$ and $\sigma_{8} = 0.9$.
The information provided by the merger tree are the virial mass $M$ of the merging subhalos and the scale factor $a$ at which they fall into the main progenitor. 
The mass resolution of the tree is $10^{8} M_{\odot}$. 
\\
Knowing the epoch in the past at which each satellite falls into the Milky Way's halo and the initial position of the merging satellite in phase-space at that time, ($\mathbf{r_{\rm{sat}}}$, $\mathbf{v_{\rm{sat}}}$), the model evolves the system analytically backwards in time. 
This allows us to map out the full velocity distribution within the ``ideal'' detector, which is located at the Solar position, at a distance ${\bf r}_{0} = (8.5, \, 0, \, 0) \,$ kpc from the Galactic centre.
We assume that each subhalo falls into the host system on a radial orbit from a random position on a sphere of radius $r_{\rm{sat}} = r_{\rm{vir}}$, where $r_{\rm{vir}}$ is the (redshift dependent) virial radius of the Milky Way halo.
Experiments reveal that these assumptions are not critical. 
For example, replacing the radial orbits with orbits that have a peri-to-apo-centre ratio of 1:6, the median found in cosmological simulations (Ghigna et al.\ 1998, Diemand et al.\ 2008), results in velocity distributions that are indistinguishable. 
Similarly, starting the disintegration of each subhalo at its first pericentre passage rather than at the Milky Way's virial radius on its initial infall does not substantively alter the final velocity distribution.
The quantity $v_{\rm{sat}}$ is approximated by the velocity of a body falling from an infinite distance, 
\begin{equation} \label{eq:vel_infinite}
v_{\rm{sat}} = \sqrt{\frac{2GM}{r_{\rm{vir}}}} \ .
\end{equation}
Finally, we use the virial theorem 
\begin{equation}  \label{eq:dispersions1}
M_{\rm{vir}} = \frac{r_{\rm{vir}} \sigma^{2}_{\rm{v}}}{G} \ , 
\end{equation}
and the definition of virial mass
\begin{equation}  \label{eq:dispersions2}
M_{\rm{vir}} = \frac{4}{3} \pi \Delta_{\rm{vir}} \rho_{\rm{crit}} r^{3}_{\rm{vir}} \ , 
\end{equation}
to relate the velocity dispersion, the spatial extent and the virial mass. 
The virial overdensity $\Delta_{\rm{vir}}$ is defined as the density relative to the mean density within $r_{\rm{vir}}$ times the critical density $\rho_{crit}$ at that redshift (Bryan \& Norman 1998), and $\rho_{\rm{crit}}$ is the critical density.
Assuming a value for the concentration parameter $c = r_{\rm{vir}} / \sigma_{\rm{s}} = 10$ (Bullock et al. 2001; Benson 2005), Eqs. (\ref{eq:dispersions1}) - (\ref{eq:dispersions2}) give an estimate of the two characteristic scales of the merging subhalo. 
\\
To summarise, the principal steps of this modelling process are:
\begin{itemize}
 \item Pick a merger tree that gives a realistic representation of a Milky Way-like system.
 \item Calculate the initial conditions for each subhalo in the merger tree.
 \item Evaluate the velocity distribution, observable within the detector for each subhalo, using the model developed in Paper I. 
 \item Evaluate the total velocity distribution by summing all the constituent mergers.
\end{itemize}

\section{Results}

The model we have developed above is not intended to predict in quantitative detail the experimental signal that a terrestrial detector would observe. 
Nevertheless, it is interesting to define quantities related to the ones measured by directional detectors, with the aim of investigating qualitatively the properties of the system. 
In Section 3.1 we first determine the local DM energy spectrum of a Milky Way-like halo as a quantity that can be directly measured by experiments. 
Secondly, we present the velocity distribution as a function of direction and speed, quantities that can give important indications about the evolution history of the Galactic halo.
In Section 3.2 we present the contribution to the velocity distribution produced by the addition of a new progenitor to the same Milky Way-like halo, in order to determine if late-merging subhalos might be detectable.

\subsection{Formation of a Milky Way-like dark matter halo}
\label{sec:section1}

%
%
\begin{figure} 
\includegraphics[height=0.25\textheight, width=0.48\textwidth]{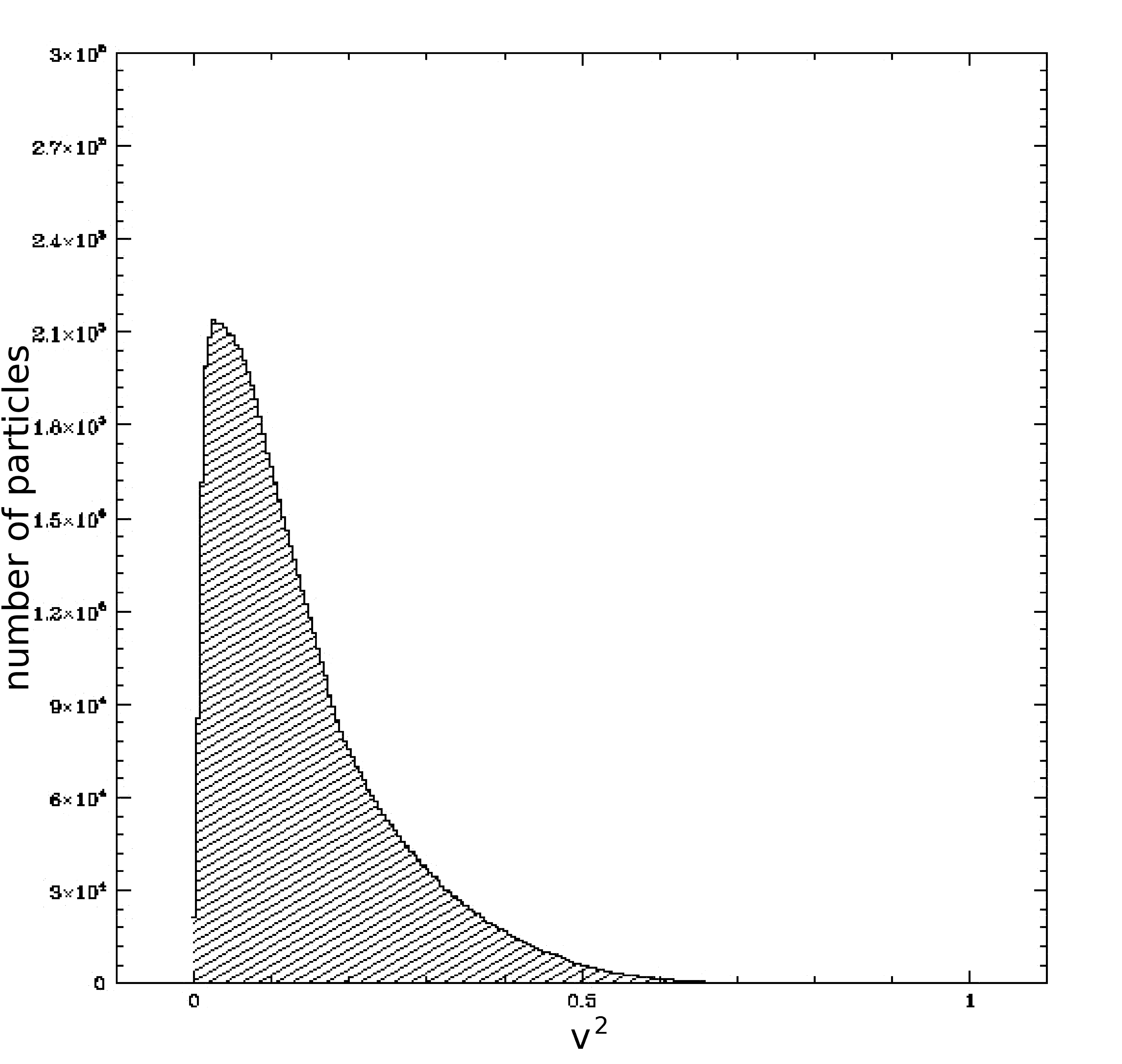}
 \caption{Energy spectrum of a Milky Way-like halo for an evolution time of $ \simeq 13.2 \,$Gyr. 
The merger history of this halo of $10^{12} M_{\odot}$ does not include any recent major merger.
One velocity-unit corresponds to $570$ km s$^{-1}$.}
\label{fig:energy_spectrum}
\end{figure}

%
%
\begin{figure} 
\includegraphics[height=0.25\textheight, width=0.48\textwidth]{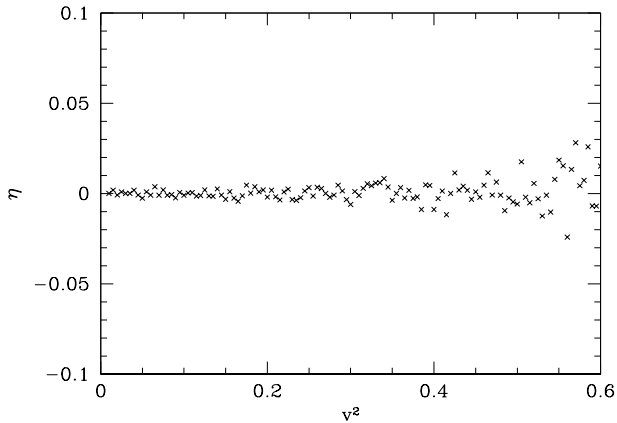}
 \caption{Fractional departure, $\eta$, of the velocity-squared of the Galactic halo simulated in Figure \ref{fig:smooth_original} from a smooth distribution for an evolution time $t \simeq 13.2$ Gyr.
As in Fig. \ref{fig:energy_spectrum}, one velocity-unit corresponds to $570$ km s$^{-1}$.}
\label{fig:df1_departure}
\end{figure}			

The energy spectrum is an observable quantity, and the analysis of its features can shed light on the evolution history of a system. 
For example, the particles of a stream produced by the recent merger of a subhalo normally group around particular values of the energy, forming ``overdensities''. 
Fig. \ref{fig:energy_spectrum} presents the energy spectrum of the Galactic halo for an evolution time of $\simeq 13.2 \,$Gyr.
In the plot we see that the energy spectrum does not show any small-scale feature. 
This smoothness arises because the halo is composed of a number of particles large enough to wash out any possible ``overdensity'' produced by single streams.  
To determine the level of residuals from a completely smooth distribution, we compare the current configuration to that obtained by evolving the system to many times the current age of the Milky Way, thus ensuring that it is fully relaxed. 
The fractional departure from this relaxed configuration is then given by:
\begin{equation} \label{eq:frac_depart}
\eta = \frac{N_{\rm{e}} - N_{\rm{rel}}}{N_{\rm{rel}}} \ ,
\end{equation}
where $N_{\rm{e}}$ is the number of particles in a range of velocities at $t_{\rm{i}}$ and $N_{\rm{rel}}$ is the number of objects in the same range once the system is relaxed.
%
%
\begin{figure*} 
\centering
\includegraphics[height=0.65\textheight,width=0.65\textwidth,angle=270]{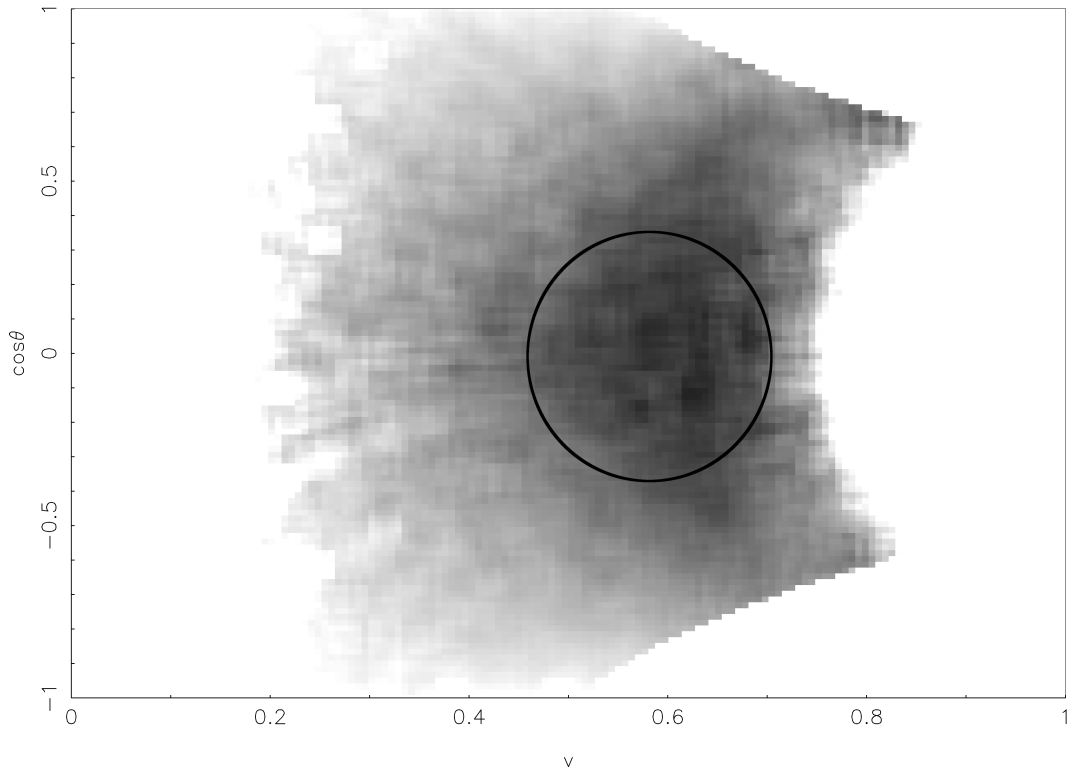}
\caption{The smoothed velocity distribution of a Milky Way-like DM halo of mass $10^{12} M_{\odot}$ as a function of $\cos \theta$ and $v$. 
The angle distribution is smoothed over a ($5\times5$) box. 
The detector is located at a position $r_{0} = (8.5, \, 0, \, 0) \,$ kpc. 
We use an array of dimensions $n_{v}, n_{\theta} = (150 \times 120)$.
The dimension of a pixel on the x-axis is $\simeq 3.8$ km s$^{-1}$. 
The angular resolution on the y-axis varies between $10^{\circ}\, 30'$ (for small angles) and $ \simeq 1^{\circ}$ (for large angles).
The greyscale, with the shade of each pixel determined by the corresponding array of the distribution function $f(\mathbf{r}, \mathbf{v})$, covers six orders of magnitude. 
Black corresponds to the upper limit and light grey to the lower one.
The black circle highlights the region where overdensities, relative to the surrounding pixels, are present.}
\label{fig:smooth_original}
\end{figure*}
%
\\
In Figure \ref{fig:df1_departure} we present $\eta$ for the same system as Fig. \ref{fig:energy_spectrum}. 
From the plot it is evident that the fractional departure is absolutely indiscernible from the distribution that the halo assumes when completely relaxed.
The absence of any feature leads to the conclusion that the ultra-fine DM distribution in the Solar neighbourhood is composed of a huge number of streams. 
These overdense structures, clearly visible when we analyse the interaction of a single satellite, overlap, generating a smooth DM distribution and consequently a featureless fractional departure.

In principle, more information might be preserved if we look at the full velocity distribution as a function of angle, as recorded by some DM detectors. 
Fig. \ref{fig:smooth_original} is a representation of the velocity distribution $f(\mathbf{r}, \mathbf{v})$ of the Milky Way-like halo simulated above in the $( \cos \theta, v)$ space, where $\theta$ is defined as the angle at which DM particles enter the detector, measured relative to the direction of Solar motion in the Milky Way.
The plot is grey-scaled, and the shade of each pixel covers a range of six orders of magnitude in $f(\mathbf{r}, \mathbf{v})$. 
This range is large enough to show the most significant contributions to the velocity distribution of the system.
Assuming $v_{\rm{max}} = 570$ km s$^{-1}$, a value slightly larger than the escape speed from the Milky Way ($v_{\rm{esc}} = 544$ km s$^{-1}$, Smith et al. 2007), and $v_{min} = 0$ km s$^{-1}$, the velocity resolution of each cell along the $x$-axis is $\simeq 3.8$ km s$^{-1}$. 
The angular resolution varies between $\simeq 10^{\circ}$ (for small angles) and $ \simeq 1^{\circ}$ (for large angles).
To investigate the presence of overdensities in the angle distribution we apply a boxcar smoothing technique. 
This removes all the small differences in the velocity-space distribution, mainly due to pixelation artifacts. 
Fig. \ref{fig:smooth_original} shows that in a system resembling our galaxy no clear features are present, except for a central overdensity. 
The overdensities are relative to the surrounding pixels. 
The fact that the overdensities are not easily distinguishable means that a full quantitative analysis is not justified, but there is clearly information in these structures.
The particles composing this region have velocities in the range $ 285 $ km s$^{-1} \lesssim v \lesssim 400$ km s$^{-1}$, which in the plot correspond to $ 0.5 \lesssim v \lesssim 0.7$, and angles $ 60^{\circ} \lesssim \theta \lesssim 120^{\circ}$ ($ -0.25 \lesssim \cos \theta \lesssim 0.25$).
A second level of overdensities seems to be present in the same region, though it is difficult to separate them from the main one.
These weak features are due to recent and incomplete merger activity, and their existence raises the possibility of disentangling at least the recent merger history of the halo. 
This could be achieved by directional detectors with angle and velocity resolution of $\simeq 1^{\circ}$ and $\simeq 3.8$ km s$^{-1}$.
\\
An investigation on the origin of these two levels of overdensities can be carried out by analysing the contributions to the final velocity distribution of progenitors with similar mass. 
Using the same smoothing technique applied above, in Figure 5 we group together the satellites with mass of the same order of magnitude.
The three groups we adopt are: $10^{8} - 10^{9}\, M_{\odot}$ (Fig. 5a), $10^{9} - 10^{10}\, M_{\odot}$ (Fig. 5b) and $10^{10} - 10^{11}\, M_{\odot}$ (Fig. 5c).
The three figures show that the contributions from these groups of subhalos are heterogeneous.
The number of direct progenitors in the range of mass $10^{8} - 10^{9}\, M_{\odot}$ is large ($\sim 10^3$) and their total contribution to the final velocity-space distribution is not negligible.
Most of the particles have $ 75^{\circ} \lesssim \theta \lesssim 105^{\circ}$ ($ -0.25 \lesssim \cos \theta \lesssim 0.25$), but there is no evidence for a preferential range of velocity.
This is due to the fact that these progenitors are compact and the timescale on which they get disrupted is longer than that of more massive systems. 
As a consequence, even after an evolution time equal to the Milky Way's age they are not completely disrupted.
\\
Looking at Fig. 5b and 5c we note that the contributions produced by subhalos of mass $10^{9-10}\, M_{\odot}$ and $10^{10-11}\, M_{\odot}$ are very similar.
A significant difference can be seen: 
the velocity distribution produced by $10^{9}\, M_{\odot}$ subhalos does not show any particular feature, while that from $10^{10}\, M_{\odot}$ satellites has a central overdensity.
This suggests that the most pronounced overdensity of Fig. \ref{fig:smooth_original} is mostly produced by satellite with masses in the range $10^{10-11}\, M_{\odot}$. 
%
%
\begin{figure*}
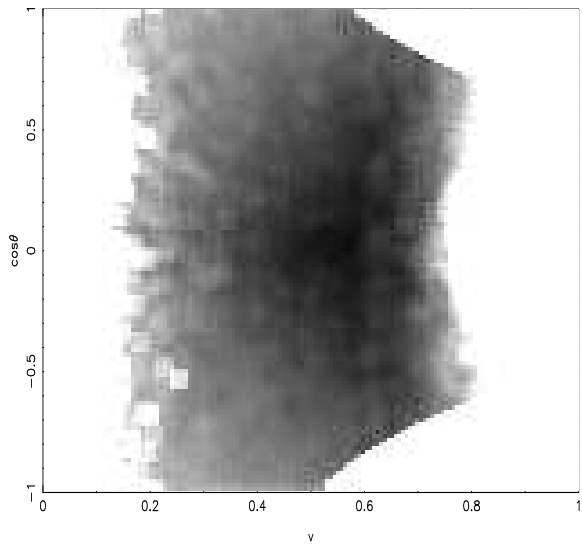
 
\centering
\begin{minipage}{0.48 \textwidth}
\centering
Fig. 5a: $10^8 \, M_{\odot} \leq M < 10^9 \, M_{\odot}$\\
\label{fig:contrib8-9} \includegraphics[height=0.35\textheight,width=0.95\textwidth,angle=270]{DF1_contribution8-9_11-17.ps}
\end{minipage}
\begin{minipage}{0.48 \textwidth}
\centering
Fig. 5b: $10^9 \, M_{\odot} \leq M < 10^{10} \, M_{\odot}$\\
\label{fig:contrib9-10} \includegraphics[height=0.35\textheight,width=0.95\textwidth,angle=270]{DF1_contribution9-10_smooth.ps}
\end{minipage}
\begin{minipage}{0.48 \textwidth}
\centering
Fig. 5c: $10^{10} \, M_{\odot} \leq M < 10^{11} \, M_{\odot}$\\
\label{fig:contrib10-11} \includegraphics[height=0.35\textheight,width=0.95\textwidth,angle=270]{DF1_contribution10-11_smooth.ps}
\end{minipage}
 \caption{Series of plots presenting the contribution to the Milky Way-like halo angle distribution, simulated in Fig. \ref{fig:smooth_original}, 
from the satellites with mass in the range $10^8 \, M_{\odot} \leq M < 10^9 \, M_{\odot}$ (Fig. 5a), $10^9 \, M_{\odot} \leq M < 10^{10} \, M_{\odot}$ (Fig. 5b), and $10^{10} \, M_{\odot} \leq M < 10^{11} \, M_{\odot}$ (Fig. 5c). 
The greyscale and the dimensions of each pixel are the same as used Fig. \ref{fig:smooth_original}.}
\label{fig:contrib}
\end{figure*}

\subsection{Contribution of single merger events to the ultra-fine distribution of a Milky Way-like system}

\begin{figure*}
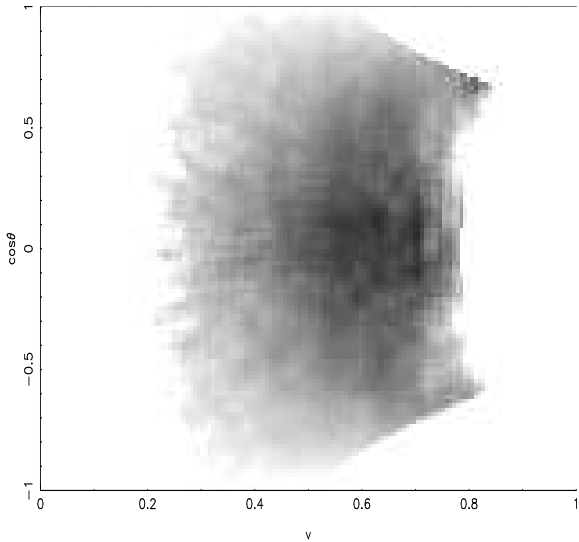
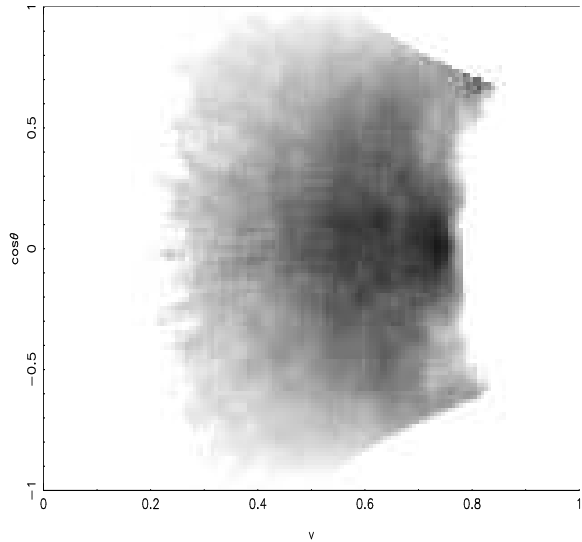

\begin{minipage}[t]{0.48 \textwidth}
\centering
Fig. 6a: $t_{1} \simeq 1.75 \, $ Gyr\\
\label{fig:df1+m10_t22} \includegraphics[height=0.35\textheight,width=0.95\textwidth,angle=270]{smooth_DF1+m10_22.ps}
\end{minipage}
\begin{minipage}[t]{0.48 \textwidth}
\centering
Fig. 6c: $t_{3} \simeq 7.5 $ Gyr\\
\label{fig:df1+m10_t52} \includegraphics[height=0.35\textheight,width=0.95\textwidth,angle=270]{smooth_DF1+m10_520.ps}
\end{minipage}
%
%
%
\begin{minipage}[t]{0.48 \textwidth}
\centering
Fig. 6b: $t_{2} \simeq 2.0 \, $ Gyr\\
\label{fig:df1+m10_t520} \includegraphics[height=0.35\textheight,width=0.95\textwidth,angle=270]{smooth_DF1+m10_52.ps}
\end{minipage}
\begin{minipage}[t]{0.48 \textwidth}
\centering
Fig. 6d: $t_{4} \simeq 13 \, $ Gyr\\
\label{fig:df1+m10_t885} \includegraphics[height=0.35\textheight,width=0.95\textwidth,angle=270]{smooth_DF1+m10_885.ps}
\end{minipage}
\caption{Series of snapshots presenting the angle-velocity plot at $z=0$ of a system composed of the Milky Way-like halo and a subhalo of mass $10^{10} \, M_{\odot}$.
The four plots describe configurations in which the progenitor falls into the host halo on a radial orbit at different times.
In Fig. \ref{fig:df1_m10}a it fell $\simeq 1.75 \, $Gyr ago. 
Fig. \ref{fig:df1_m10}b corresponds to a infall time of $\simeq 2.0 \, $Gyr ago, while in Fig. \ref{fig:df1_m10}c $t_3 \simeq 7.5 \, $Gyr ago.  
Finally, in Fig. \ref{fig:df1_m10}d the satellite merged $\simeq 13 \, $Gyr ago.
The greyscale and the dimensions of each pixel of the four snapshots are the same as Fig. \ref{fig:smooth_original}, and the angle distribution is smoothed over a ($5\times5$) box.}
\label{fig:df1_m10}
\end{figure*}

This motivates the study of the contribution from single mergers. 
The method we adopt to investigate this issue consists of adding a DM satellite to the Galactic halo and of analysing its contribution to the final velocity-space distribution.
The properties of the additional progenitor that we vary are the mass and the time at which it falls into the Galaxy.
For the mass, we consider four different cases: $10^{7} \, M_{\odot}$,  $10^{8} \, M_{\odot}$, $10^{9} \, M_{\odot}$ and $ 10^{10} \, M_{\odot}$.
The main reason that motivates the limit of  $ 10^{10} \, M_{\odot}$ is the absence of major mergers in the recent history of the Milky Way, based on theoretical considerations which show that the accretion of a $1:10$ satellite galaxy could destroy the disk of the Milky Way (Purcell et al. 2009). 
To get a complete overview of the evolution of the mergers we select four times, looking back into the past, at which the progenitor fell into the host halo:
\begin{itemize}
 \item $t_{1} \simeq 1.75 \, $Gyr ago ($z=0.14$): the subhalo is making its first passage through the detector today at redshift $z=0$.
 \item $t_{2} \simeq 2.0 \, $Gyr ago ($z=0.17$): stage at which the satellite is not passing through the detector today. 
 \item $t_{3} \simeq 7.5 \, $Gyr ago ($z=1.0$): the merger is at an intermediate stage of its evolution.
 \item $t_{4} \simeq 13 \, $Gyr ago ($z=8.5$): the satellite merged at an early stage of the formation of the Milky Way.
\end{itemize}
%
%
\begin{table}
\centering
    \begin{tabular}{| l || c | c | c | c | }
\hline
\, Mass $M_{\odot}$ / Infall Time (Gyr ago) & {1.75} & {2.0} & {7.5} & {13} \\
\hline
\hline
\, $10^{7}$ & $\times$ & $\times$ & $\times$ & $\times$  \\
\hline
\, $10^{8}$ & $\times$ & $\times$ & $\times$ & $\times$ \\
\hline
\, $10^{9}$ & \checkmark & $\times$ & \checkmark & \checkmark \\
\hline
\, $10^{10}$ & \checkmark & \checkmark & \checkmark & \checkmark \\
\hline
    \end{tabular}
  \caption{Table showing the presence (or absence) of differences between the velocity distribution of a system composed of the Milky Way-like halo plus the contribution of a single subhalo and the Milky Way-like halo only. 
Different masses are considered for the new progenitor, as well as different infall times. 
The symbol $\checkmark$ is used if there are differences in these two values of $f(\mathbf{r},\mathbf{v})$ (in a range of six orders of magnitude).
On the other hand, if the values of the two $f(\mathbf{r},\mathbf{v})$ are identical, the symbol $\times$ is used.}
\label{tab:merger_orig}
\end{table}
The results of these simulations are summarised in Table \ref{tab:merger_orig}, which shows the comparison between the velocity distribution of a system composed of the Milky Way-like halo plus the contribution at $z = 0$ of one of the extra progenitors and the Milky Way-like halo only.
\\
The effect of the addition of an extra progenitor of mass between $10^{7} \, M_{\odot}$ and $10^{8} \, M_{\odot}$ to the merger tree of the Galactic halo is always negligible. 
The contribution is not negligible anymore when the mass of the new merging subhalo is equal to $10^{9} \, M_{\odot}$. 
If plotted separately, without the presence of the ``background'' distribution produced by the Galactic halo, the contribution is present and clearly visible. 
Unfortunately, once we merge in the same plot these two contributions, the one of the extra progenitor is not visible anymore because it is concealed by the smooth ``background'' distribution produced by the Galactic halo.
\\
The only case in which the merger of an extra progenitor generates clear features in the angle distribution of the Milky Way is when the mass of the subhalo is equal to $10^{10} \, M_{\odot}$. 
This configuration is shown in the four snapshots of Fig. \ref{fig:df1_m10}. 
Fig. \ref{fig:df1_m10}a represents the fall of the subhalo into the Galaxy approximately 1.75 Gyr ago.
Two vertical, narrow stripes are clearly visible. 
They both have an angular width of approximately $ 15^{\circ}-20^{\circ}$: 
the first does not have sharp contours, while the second is centred in the range $79^{\circ} \lesssim \theta \lesssim 96^{\circ}$ ($-0.2 \lesssim \cos \theta \lesssim 0.2$).  
The presence of these two stripes shows that the subhalo is beginning to get stripped and that streams start to form. 
Their width allows them to be potentially detectable by current directional experiments.
\\
At $t_{2}$ the two vertical stripes have almost completely disappeared (Fig. \ref{fig:df1_m10}b). 
This behaviour is expected because the main body of the satellite is still coherent and it is not located anywhere near the Solar neighbourhood. 

At $t_{3}$ the satellite has performed several orbits around the Galactic centre (Fig. \ref{fig:df1_m10}c). 
At each orbit the phase mixing acts on the satellite, increasing the number of streams and the amount of debris into which the subhalo is disrupted. 
The streams overlap and the velocity distribution of the subhalo becomes smooth and uniform.
Nevertheless, there is a narrow, vertical stripe at velocity $\simeq 430$ km s$^{-1}$ ($v \simeq 0.75$), centred in the range $79^{\circ} \lesssim \theta \lesssim 96^{\circ}$ ($-0.2 \lesssim \cos \theta \lesssim 0.2$). 
This is due to the presence of a dense, massive stream at the Solar position, and it suggests that the disruption of the $10^{10} \, M_{\odot}$ satellite is not complete. 
\\
%
%
Finally, Fig. \ref{fig:df1_m10}d presents the configuration in which the merger of the subhalo happened 13 Gyr ago.
The stripes that are present in other plots are now completely washed out.
The absence of stripes confirms the presence of an almost homogeneous distribution. 
Nevertheless, the remnant of the central part of the satellite is not completely disrupted and it still shows up as an overdensity. 
This overdensity is located in the speed range $400$ km s$^{-1}$ $ \lesssim v \lesssim $ $430$ km s$^{-1}$ ($0.7$ km s$^{-1}$ $ \lesssim v \lesssim $ $0.75$ km s$^{-1}$), with an angular width of about $\simeq 5^{\circ}$ ($-0.05 \lesssim \cos \theta \lesssim 0.05$). 

\section{Discussion}

The conclusion inferred from the analysis of the energy spectrum of a system resembling the Milky Way is that the ultra-fine DM distribution in the Solar neighbourhood is composed of a large number of streams.
The streams overlap and produce a smooth distribution. 
Our result is in agreement with the conclusion of VW, which has been reached using an approach complementary to this analysis. 
VW's technique must be used in tandem with a N-body simulation. 
This implies the need of a large computational time and it requires a softening length of $3.4$ kpc to achieve stable results.
In our model, the gravitational force does not have to be artificially softened. 
Consequently, the phase-space can be explored very rapidly and accurately in a single timestep, without the computational overhead of numerical integration, and the mpc-scale resolution required by direct detection experiments can be reached by using a less realistic gravitational potential. 
The fact that these very different techniques give the same answer provides some confidence that the results are not an artifact of the approximations made in each case.   
\\
When we investigate the velocity distribution $f(\mathbf{r}, \mathbf{v})$ in the $(\cos \theta, v)$ space, the results suggest the presence of two levels of overdensities.
They are the marks left by recent merger activity and we have verified that they have been mostly produced by satellites with mass in the range $10^{10-11}\, M_{\odot}$. 
The probability of detecting such overdensities is however low.
This is mainly caused by both the lack of strong features, and the limited angular resolution of the current generation of instruments.
Nowadays this resolution is $\simeq 15^{\circ}$  (Dujmic et al. 2008, Ahlen et al. 2010), one order of magnitude larger than that required for the detection of these overdensities ($\simeq 1^{\circ}$). 

The analysis of the contribution from single mergers to the ultra-fine DM distribution of a Milky Way-like halo shows the presence of features when the mass of the additional satellite is equal or larger than $10^{9} \, M_{\odot}$. 
The features generated by the merger of a satellite of mass $\leq 10^{9} \, M_{\odot}$ are concealed by the smooth ``background'' distribution produced by the Galactic halo, that dominates the $(\cos \theta, v)$ space.
Once this ``background'' is subtracted, it is possible to see the imprint left by the additional satellite.
Nevertheless, its angular width is too small ($\simeq 1^{\circ}$) to be detectable with the present-day technology.
Moreover, the angular width of the features generated by the merger of a $10^{10} \, M_{\odot}$ progenitor is $ 15^{\circ}-20^{\circ}$, which is accessible to ongoing experiments.
This raises the possibility to detect these peculiar features. 
This possibility depends on the probability of finding the remnant of a $10^{10} \, M_{\odot}$ subhalo in the Solar neighbourhood and that this subhalo merged into the Galactic halo in the recent past ($\lesssim$ 7.5 Gyr ago). 
This probability is not negligible: various works (Stewart et al. 2008, Purcell et al. 2007, Zentner 2007) found that the mass of a DM halo, $M_h$, is provided mostly by the accretion of subhalos of mass $\sim (0.05-0.15) M_h$.
Furthermore, it has been estimated that the probability of a Milky Way-like system of having experienced a merger with a subhalo of $10^{11} M_{\odot}$ in the last 8 Gyr is $\sim 80 \%$ (Stewart et al. 2008).

\section{Summary}

In this paper we have refined the method developed in Fantin et al. (2008), applying it to the formation of the Milky Way in a cosmological context.
The refinement consists of combining a merger tree, which describes the hierarchical growth of a Milky Way-like halo, with the model previously developed, which simulates the interaction between an unbound system of particles and a larger parent halo. 
This allows us to produce a more complete and realistic treatment of the merger history of the Galactic halo, mapping out the sub-mpc-scale structure of the halo. 
Using such a simulation we can obtain new insights into the likely signature of a halo merger event in a small terrestrial DM detector.
\\
We find that the velocity distribution does not contain any clear feature.
This is caused by the overlapping of a huge number of streams, which generates a uniform and smooth velocity distribution.
This result is in agreement with recent works, such as Vogelsberger \& White (2011) and Schneider et al. (2010).
Detectable signatures in the ultra-fine velocity distribution in the Solar neighbourhood are present only in the case of the recent merger into the Galactic halo of a subhalo of $10^{10} \, M_{\odot}$. 
In this particular configuration the angular width of these features is larger than $10^{\circ}$, a resolution accessible to ongoing experiments. 
Finally, in the scenarios that we have analysed, the presence of some overdensities has been identified. 
The current generation of detectors does not have the angular resolution required to observe these features, but a future generation of detectors with resolution of $\sim 1^{\circ}$ would resolve them, allowing the recent history of the Milky Way to be revealed. 

\section{Acknowledgements}
We thank Andrew Benson for providing the merger trees used to develop the project, and Frazer Pearce for useful discussions.

\label{lastpage}

\end{document}